\documentclass[twocolumn]{aastex631}

\usepackage{soul}
\usepackage{amsmath}
\usepackage{listings}

\begin{document}

\title{Back to the Starting Point: on the Simulation of Initial Magnetic Fields and Spin Periods of Non-accretion Pulsars}

\author[0000-0002-9739-8929]{Kun Xu}
\affiliation{School of Astronomy and Space Sciences, University of Chinese Academy of Sciences, Beijing, People’s Republic of China; xukun@smail.nju.edu.cn}
\affiliation{Key Laboratory of Optical Astronomy, National Astronomical Observatories, Chinese Academy of Sciences, Beijing, People’s Republic of China}

\author[0000-0001-5532-4465]{Hao-Ran Yang}
\affiliation{Department of Astronomy, Nanjing University, Nanjing 210023, People’s Republic of China}
\affiliation{Key Laboratory of Modern Astronomy and Astrophysics, Nanjing University,
Ministry of Education, Nanjing 210023, People’s Republic of China}

\author[0000-0001-8356-2233]{Ying-Han Mao}
\affiliation{Department of Astronomy, Nanjing University, Nanjing 210023, People’s Republic of China}
\affiliation{Key Laboratory of Modern Astronomy and Astrophysics, Nanjing University,
Ministry of Education, Nanjing 210023, People’s Republic of China}

\author[0000-0001-9565-9462]{Xiao-Tian Xu}
\affiliation{Argelander-Institut f\"ur Astronomie, Universit\"at Bonn, Auf dem H\"ugel 71, 53121 Bonn, Germany; xxu@astro.uni-bonn.de}

\author[0000-0002-0584-8145]{Xiang-Dong Li}
\affiliation{Department of Astronomy, Nanjing University, Nanjing 210023, People’s Republic of China}
\affiliation{Key Laboratory of Modern Astronomy and Astrophysics, Nanjing University,
Ministry of Education, Nanjing 210023, People’s Republic of China}

\author{Jifeng Liu}
\affiliation{School of Astronomy and Space Sciences, University of Chinese Academy of Sciences, Beijing, People’s Republic of China; xukun@smail.nju.edu.cn}
\affiliation{Key Laboratory of Optical Astronomy, National Astronomical Observatories, Chinese Academy of Sciences, Beijing, People’s Republic of China}
\affiliation{WHU-NAOC Joint Center for Astronomy, Wuhan University, Wuhan, People’s Republic of China}

\begin{abstract}

Neutron stars (NSs) play essential roles in modern astrophysics. Magnetic fields and spin periods of newborn (zero age) NSs have large impact on the further evolution of NSs, which are however poorly explored in observation due to the difficulty of finding newborn NSs. In this work, we aim to infer the magnetic fields and spin periods ($B_{\rm i}$ and $P_{\rm i}$) of zero-age NSs from the observed properties of NS population. 
We select non-accretion NSs (NANSs) whose evolution is solely determined by magnetic dipole radiation. We find that both $B_{\rm i}$ and $P_{\rm i}$ can be described by log-normal distribution and the fitting sensitively depends on our parameters.

\end{abstract}

\keywords{Neutron stars (1108) --- Magnetic fields (994) --- Accretion (14) --- Stellar rotation (1629)}

\section{Introduction} \label{sec:intro}

%\xiaotian{We need a definition of 'initial'. How do you define your zero-age state? }
The initial properties of neutron stars after birth are essential to understand some astrophysical problems, especially some high-energy phenomena, such as supernovae (SNe), X-ray bursts (XRBs), gamma-ray bursts (GRBs) and fast radio bursts (FRBs).
But it is hard to directly detect them in observation because the radiation from a newborn NS is obscured by the dense material ejected from supernovae \citep[e.g.][]{Igoshev2022}.
So it is still a puzzle since we know little about two basic properties of NSs: the initial magnetic fields $B_{\rm i}$ and spin periods $P_{\rm i}$ \citep[e.g.][]{Popov2010,Makarenko2021,Igoshev2022}.
%\xiaotian{Is there any other work trying to do similar thing?}

\begin{deluxetable*}{ccc}
\tablewidth{0pt}
\tablecaption{The $B_{\rm i}$ and $P_{\rm i}$ distribution from literature. \label{table:BP_refs}} 
\tablehead{\colhead{$B_{\rm i}$} & \colhead{$P_{\rm i}$} & \colhead{References}}
\startdata
$\langle {\rm log} (B_{\rm i}/{\rm G}) \rangle \sim 12.95$, $\sigma_{\rm log ~ B_i} \sim 0.55$ & $\langle P_i \rangle \sim 300$ ms, $\sigma_{P_i} \sim 0.15$ s & 1, 2 \\
$\langle {\rm log} (B_{\rm i}/{\rm G}) \rangle \sim 13.25$, $\sigma _{\rm log ~ B_i} \sim 0.6$ & $\langle P_{\rm i} \rangle \sim 0.25$ s, $\sigma_{\rm P_i} \sim 0.1 $ s & 3  \\
$\langle {\rm log} (B_{\rm i}/{\rm G}) \rangle \sim 13.0-13.2$, $\sigma _{\rm B_i} \sim 0.6-0.7$ & $P_{\rm i} \lesssim 0.5$ s & 4 \\
- & $\langle P_{\rm i} \rangle \sim 0.1$ s, $\sigma_{\rm P_i} \sim 0.1$ s & 5 \\
$\langle {\rm log} (B_{\rm i}/{\rm G}) \rangle \sim 12.44$, $\sigma _{\rm log ~ B_i} \sim 0.44$ & $\langle {\rm log} (P_{\rm i}/{\rm s}) \rangle \sim -1.04$, $\sigma_{\rm log P_i} \sim 0.53$ & 6 \\
\hline
\enddata
\tablereferences{(1) \citet{Faucher-Giguere2006}; (2) \citet{Igoshev2013}; (3) \citet{Popov2010}; (4) \citet{Gullon2014}; (5) \citet{Popov2012}; (6) \citet{Igoshev2022}.}
\end{deluxetable*}

On the other hand, the population synthesis in theory is harassed to simulate the initial properties because there are a few channels to form NSs and the mechanisms are complicated \citep{Faucher-Giguere2006,Popov2010,Gullon2014}.
Assuming a constant magnetic field to simulate the Galactic pulsars, the results can be roughly described using the log-normal distribution with mean $\langle {\rm log} (B_{\rm i}/{\rm G}) \rangle \sim 12.65$ \footnote{This is the value of the magnetic field at the equator, and the surface mean value is $\langle {\rm log} (B_{\rm i}/{\rm G}) \rangle \sim 12.95$ \citep{Popov2010}.} 
and the standard deviation $\sigma_{\rm log ~ B_i} \sim 0.55$ for the initial magnetic fields, as well as the normal distribution with mean $\langle P_{\rm i} \rangle \sim 0.3 {\rm ~s}$ and the standard deviation $\sigma_{\rm P_i} \sim 0.15 {\rm ~s}$ for the initial spin periods \citep{Faucher-Giguere2006,Igoshev2013}.
Considering magneto-thermal evolution, the optimal model gives the initial distributions $\langle {\rm log} (B_{\rm i}/{\rm G}) \rangle \sim 13.25$, $\sigma _{\rm log ~ B_i} \sim 0.6$ and $\langle P_{\rm i} \rangle \sim 0.25 {\rm ~s}$, $\sigma_{\rm P_i} \sim 0.1 {\rm ~s}$ \citep{Popov2010}.
While \cite{Gullon2014} finds that $\langle {\rm log} (B_{\rm i}/{\rm G}) \rangle \sim 13.0-13.2$, $\sigma _{\rm B_i} \sim 0.6-0.7$ and any broad distribution with $P_{\rm i} \lesssim 0.5 {\rm ~s}$ can match the data, but it depends strongly on parameters.
\cite{Popov2012} gets a Gaussian distribution with $\langle P_{\rm i} \rangle \sim 0.1 {\rm ~s}$ and $\sigma_{\rm P_i} \sim 0.1 {\rm ~s}$ by studying young samples, which are NSs associated with supernova remnants (SNRs).
While \cite{Igoshev2022} shows that the normal distribution is not consistent but a log-normal one fits well with $\langle {\rm log} (P_{\rm i}/{\rm s}) \rangle \sim -1.04$, $\sigma_{\rm log ~ P_i} \sim 0.53$ and $\langle {\rm log} (B_{\rm i}/{\rm G}) \rangle \sim 12.44$, $\sigma _{\rm log ~ B_i} \sim 0.44$ by using the similar group of NSs.
We show these results in Table \ref{table:BP_refs}.
%\xiaotian{I am confused here. I do not see how these different works motivate you to study isolated NSs. }

In this work, we aim to simulate the initial distribution of magnetic fields and spin periods by paying attention to the evolution experience of NSs after born.
This is much easier than the population synthesis which need to consider the diversity of formation mechanisms. %\xiaotian{why does this is a good assumption?}.

This paper is organized as follows. We describe the data and set up the model in Section 2 and exhibit the results in Section 3. The discussion and summary are in Section 4.

\section{The Data and Model} \label{sec:data_model}

\subsection{The data}

%\xiaotian{'non-accretion' do you mean both isolated NS and NS binaries not yet undergoing mass transfer?} 
%pulsars and thus the accreting or accreted pulsars are excluded
%\xiaotian{'accreted' do you mean post-accretion? Then how do you know whether there is a past mass transfer episode or not?}.
We make the following assumptions in order to select a sample of NSs from the ATNF pulsar catalogue\footnote{\href{http://www.atnf.csiro.au/research/pulsar/psrcat}{http://www.atnf.csiro.au/research/pulsar/psrcat}} \citep{Manchester2005} as large as possible:
\begin{enumerate}
    \item All the NSs can be simply divided into two groups: the accreting or accreted ones and the non-accretion ones;
    \item The non-accretion NSs spin down only dominated by magnetic dipole radiation;
    \item The magnetic fields of all the NANSs decay in a similar way, which can be described in a simple mathematical form, i.e., all non-standard mechanism of formation and evolution of the magnetic fields are neglected such as the re-emergence.
\end{enumerate}
However in practice, it is still very hard to select NANSs using the observed spin period and period derivative as well as the deduced magnetic field and spin-down age.

\begin{figure*}[ht!]
\plotone{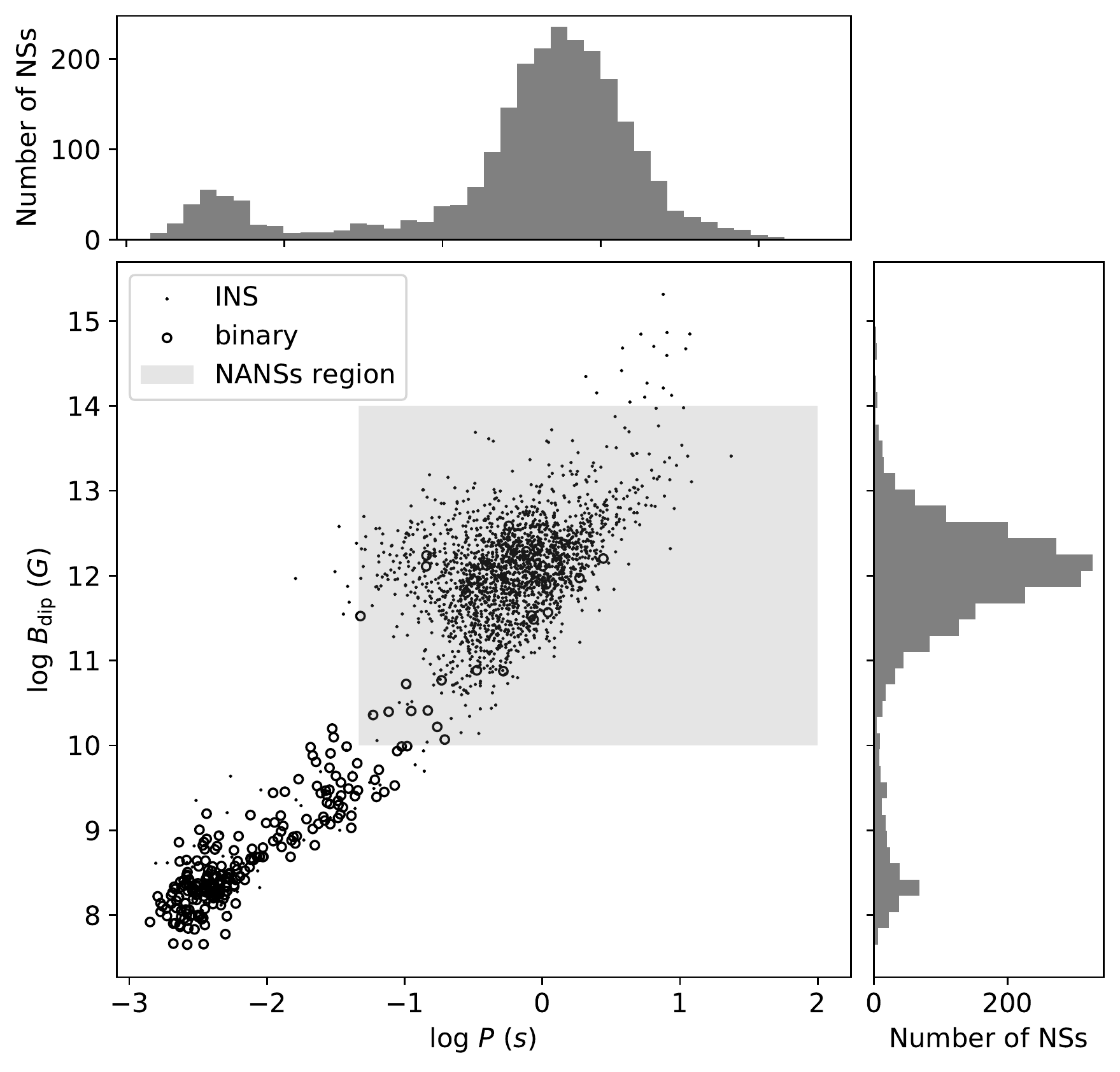}
\caption{The $B-P$ diagram of ATNF pulsars with histogram of the surface dipole magnetic fields and spin periods in the right and bottom. The dots and circles indicate the isolated neutron stars (INSs) and the neutron stars in binary systems, respectively. The dots in the grey region indicate the NANSs we selected.
\label{fig:BP_hist}}
\end{figure*}

Figure \ref{fig:BP_hist} shows the $B-P$ diagram of ATNF pulsars, where the dots and circles indicate the isolated neutron stars (INSs) and the neutron stars in binary systems, respectively.
The histograms of the surface dipole magnetic fields and spin periods are also displayed in the right and bottom, both of which exhibit bimodal features.
It shows that most of the isolated pulsars located in the upper right of the $B-P$ diagram, which are the major contributors of the main peaks in both the histograms, while most of the pulsars in binaries located in the lower left contributing mainly to the sub-peaks.
It gives an inspiration that the sub- and main peaks are dominated by accreting or accreted neutron stars and NANSs, respectively.
So one can select NANS candidates by setting lower limit of $B$ and $P$.
For simplicity, we assume that all the pulsars in binaries and the isolated pulsars with spin period smaller than $0.03 {\rm ~s}$ or surface dipole magnetic field strength smaller than $10^{10} {\rm ~G}$ maybe suffered or are suffering accretion, i.e., the isolated pulsars with $P>0.03 {\rm ~s}$ and $B>10^{10} {\rm ~G}$ belong to the non-accretion candidate groups.
In addition, since the highly magnetized NSs, especially the magnetars, express specificities in observation \citep[e.g.][]{DeLuca2006,Olausen2014,Xu2019}, which indicates that they may belong to a special group, we rule out them by setting an upper limit of $10^{14} {\rm ~G}$ for the magnetic fields, then we get the NANSs candidate sample which is studied in this paper\footnote{As described in above, it is very hard to select NANSs using the observed data, so we study the NANS candidate sample as a substitute.}.
The magnetic fields and spin periods of NANSs both favour log-normal distributions with  $\langle {\rm log} (B_{\rm i}/{\rm G}) \rangle \sim 12.04$, $\sigma _{\rm log ~ B_i} \sim 0.6$ and $\langle {\rm log} (P_{\rm i}/{\rm s}) \rangle \sim -0.22$, $\sigma_{\rm log P_i} \sim 0.4$, which are shown by grey histogram bars in Figure \ref{fig:Bc}.

\subsection{The Model}

Since it is assumed that the NANSs have not been exerted by other torques except for the dipole magnetic radiation after birth, we are likely to trace back to the starting point of these pulsars and obtain the general distribution for their initial spin periods and magnetic fields in a simple way. 

The dipole radiation could be simply written as:
\begin{equation}
    N_{\rm B}=-\frac{2\mu^2\Omega^3}{3c^3}, \label{nb}
\end{equation}
where $\Omega$, $\mu = B(t)R^3_{\rm NS}$ represent the angular velocity and the magnetic dipole moment of the NS, and $c$ is the speed of light. If the radiation loss is the only torque and then the spin evolution of the NS could be determined by the basic equation $I\dot{\Omega}=N_{\rm B}$,  in which $I = 10^{45} {\rm ~g ~cm^2}$ is the moment of inertia of an NS with typical mass $M=1.4 M_{\odot}$ and assumed to be time-independent. By integrating the both side of the equation simultaneously, we can easily derive the relation between the spin period $P$ at the time $t$ and the initial, that is,
\begin{equation}
    P^2-P^2_{\rm i}=2C_1 \times \int_{t_0}^{t} B^2(t)\ dt, \label{equ:p}
\end{equation}
where we have pack all the constant terms into the coefficient $C_1$:
\begin{equation}
    C_1=\frac{8\pi^2 R^6_{\rm NS}}{3c^3I}. \label{c1}
\end{equation}
Therefore, we can calculate the initial spin period $P_{\rm i}$ once the exact form of magnetic field evolution is given.

The magnetic fields of NSs are thought to decay under the effects of Ohmic diffusion, Hall drift, Joule heating, and so on\citep[e.g.][]{Goldreich1992,Pons2007,Konar2017}.
There are three categories of the forms of non-accretion NSs' magnetic field evolution from literature, the exponential (EXP) form \citep[e.g.][]{Usov1988,Livio1998,Gonthier2002,Gonthier2004,Kiel2008,Oslowski2011}, the power law (PL) form \citep[e.g.][]{Fu2012,Xu2022} and the composite form combined the former two \citep[e.g.][]{Aguilera2008,Zhang2012,Dall'Osso2012,Jawor2022}.
In fact, these categories can be got by following the analytical expression of magnetic field evolution introduced by \cite{Colpi2000} 
\begin{equation}
    \frac{d B(t)}{dt} = -A B(t)^{1+\alpha},
\end{equation}
where $A$ and $\alpha$ are the model parameters. Then one can get the exponential form with $\alpha=0$ \citep{Dall'Osso2012,Jawor2022}
\begin{equation}
B(t)= B_{\rm i} e^{-t/\tau_{\rm d,i}} + B_{\rm m},
\end{equation}
and the power law form with $\alpha \neq 0$
\begin{equation}
B(t)= B_{\rm i} (1+\alpha t/\tau_{\rm d,i})^{-1/\alpha} + B_{\rm m}. \label{equ:Bt}
\end{equation}
\iffalse
\begin{equation}
B(t)= \left\{
    \begin{split}
        & B_{\rm i} (1+\alpha t/\tau_{\rm d,i})^{-1/\alpha} + B_{\rm m} ~~~ (\alpha \neq 0), \\
        & B_{\rm i} e^{-t/\tau_{\rm d,i}} + B_{\rm m} ~~~~~~~~~~~~~~~ (\alpha = 0),
    \end{split} 
    \right. \label{equ:Bt}
\end{equation}
\fi
$B_{\rm i}$ and $B_{\rm m}$ are the initial and minimal magnetic fields of an NS, respectively. $\tau_{\rm d,i}=\tau_{\rm d}/(B_{\rm i}/10^{15} {\rm ~G})^{\alpha}$, where $\tau_{\rm d}$ is the field-decay timescale. $\alpha$ is the decisive parameter to the form, which is relative to the decaying rate of the magnetic field.

\begin{figure}[ht!]
\plotone{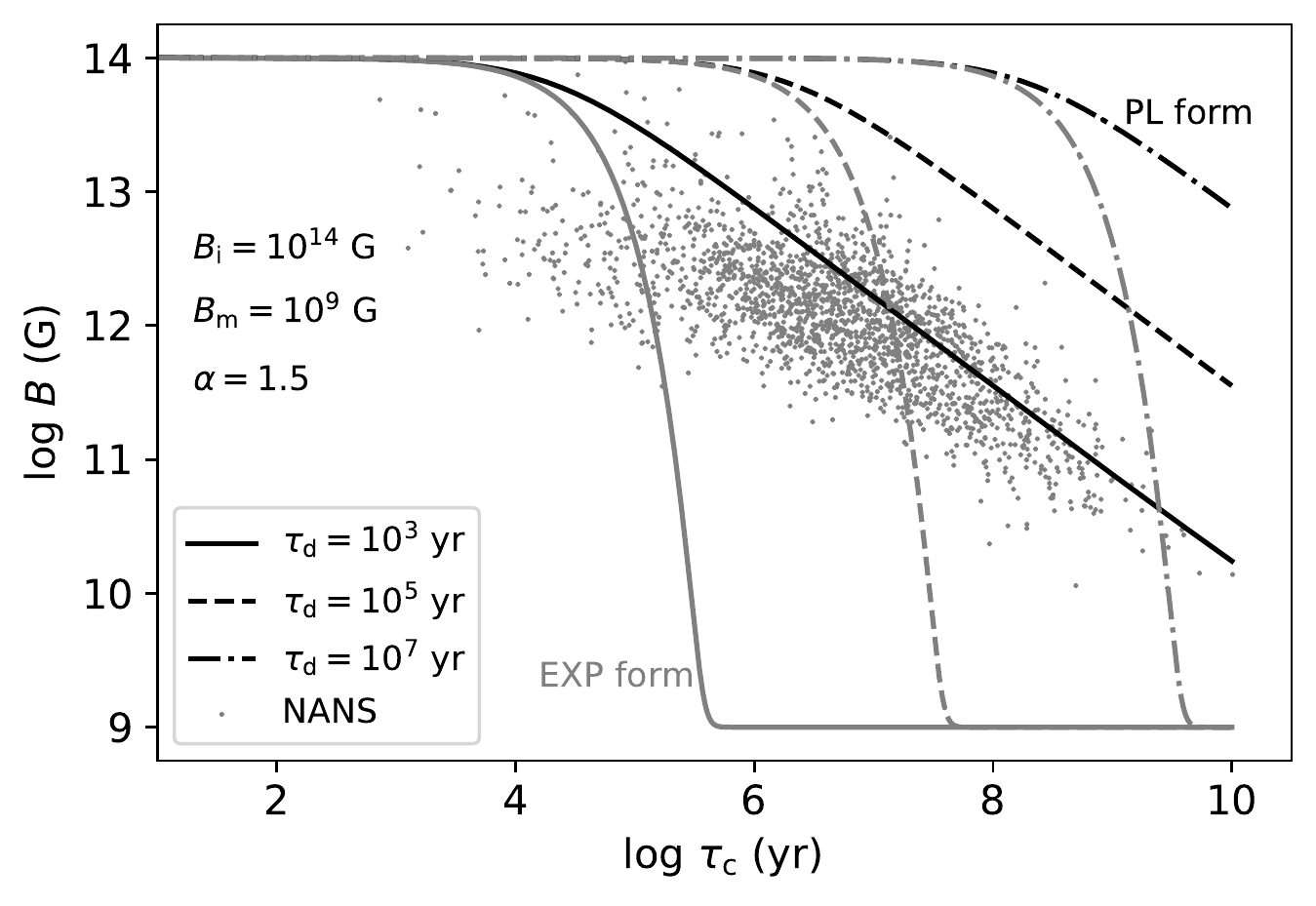}
\plotone{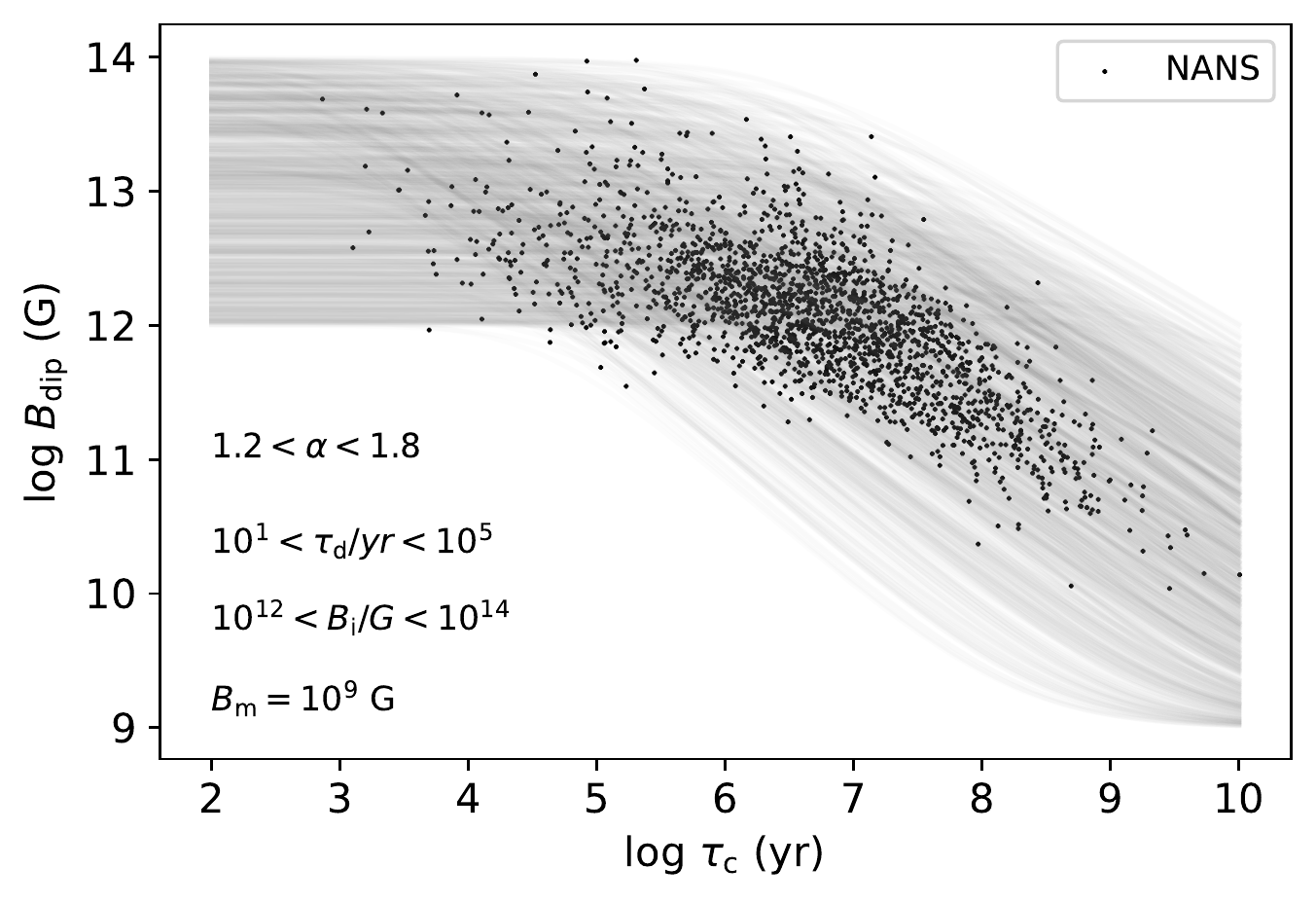}
\caption{The surface dipole magnetic field strength $B_{\rm dip}$ versus the characteristic age $\tau_{\rm c}$ of non-accretion neutron stars (the dots). 
Upper panel: the grey and black lines display the the exponential and power law form of NANSs' magnetic field evolution tracks with $\tau_{\rm d}=10^3 {\rm ~yr}$ (solid lines), $10^5  {\rm ~yr}$ (dashed lines), $10^7 {\rm ~yr}$ (dot-dashed lines), respectively.
It shows that the power law form with adjustable parameters can fit well the $B_{\rm dip}-\tau_{\rm c}$ trend of NANSs.
Lower panel: the grey lines indicate the power law form with different parameters which are listed in the diagram.
\label{fig:Btau_pl}}
\end{figure}

We compare these categories and find that the lines of the exponential form can not cover the NANS dots well because they decay too fast \footnote{Maybe it can explain some NSs with magnetic fields rapidly decaying \citep[e.g.][]{Mendes2018,Igoshev2021}.}, while the power law form is favoured by the dots trend and the lines can cover all the dots with feasible parameter space (the upper panel in Figure \ref{fig:Btau_pl}).

Although the favoured form of magnetic field evolution is got, we cannot ascertain the age of NSs since it is hard to detect its growth ring in observation. 
But in theory, one can calculate its characteristic (or spin-down) age $\tau_{\rm c} = P/2\dot{P}$, which is the only age can be easily derived for NSs.
We assume that there is an unknown relation between the physical age $\tau$ and $\tau_{\rm c}$ of NSs and we can express as $\tau = \xi \tau_{\rm c}$, where parameter $\xi$ indicates the relation.
Then we can get the the initial magnetic fields \footnote{In general, $B_{\rm m}$ is much smaller than $B_{\rm i}$, so we ignore the terms including it here and in the equation of $P_{\rm i}$}
\begin{equation}
    B_{\rm i} \approx \left( B_{\rm dip}^{-\alpha}(\tau) - \frac{\alpha t}{\tau_{\rm d} (10^{15} {\rm ~G})^{\alpha}} \right)^{-1/\alpha},
\end{equation}
using the surface dipole magnetic field strength $B_{\rm dip} = 3.2 \times 10^{19} (P\dot{P})^{1/2} {\rm ~G}$, as well as the initial spin period

\begin{equation}
        P_{\rm i} \approx \left\{ P^2 - 2C_1 \times \frac{\tau_{\rm d,i} B_{\rm i}^2}{\alpha-2} \times \left[ (1+\alpha \tau / \tau_{\rm d,i})^{-\frac{2}{\alpha}+1} - 1 \right] \right\}^{1/2}.
        \label{equ:p_i}
\end{equation}

\section{Results} \label{sec:results}

%\xiaotian{before presenting your population study, it is necessary to provide an evolution example of your model, like P evolution with different input parameters. This can tell readers how your model works}

\subsection{The model without magnetic field decay}

\begin{figure}[ht!]
%\plottwo{Bc_Bi.pdf}{Bc_Pi.pdf}
\plotone{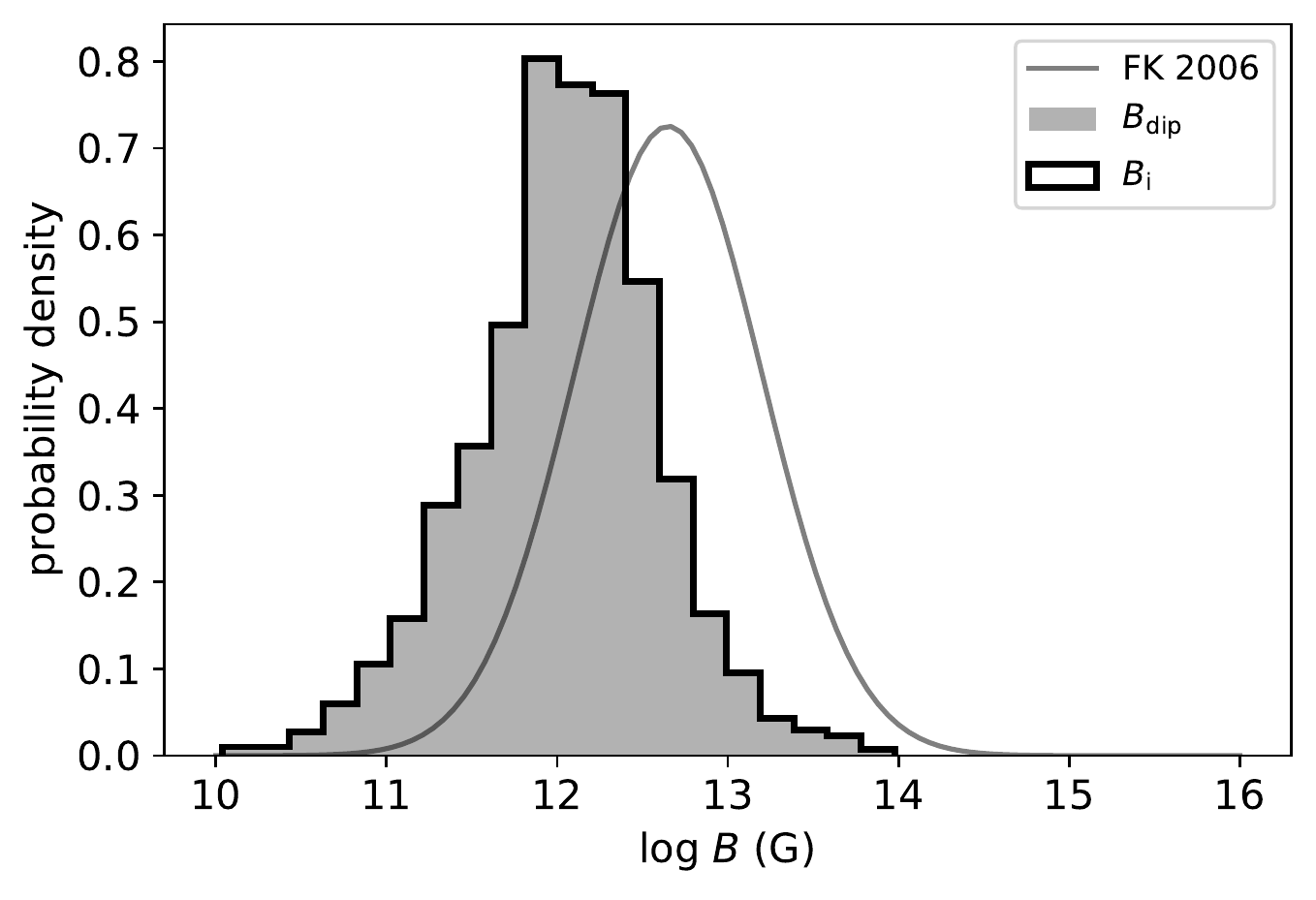}
\plotone{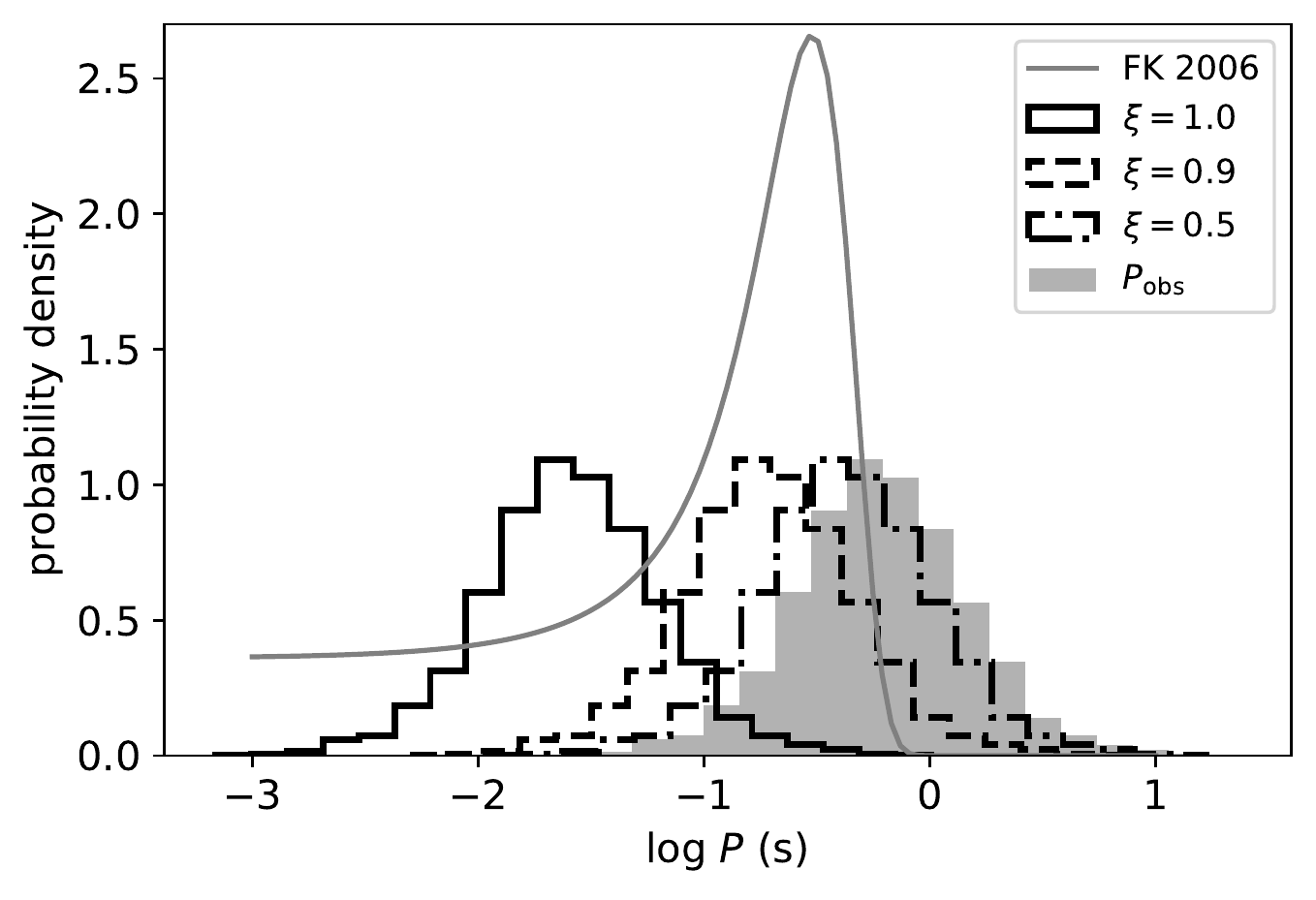}
\caption{The distribution of magnetic field (upper panel) and spin period (lower panel).
The grey histogram bars and the grey line indicate that of the magnetic field $B_{\rm dip}$ (the spin period $P_{\rm obs}$) of the NANSs and the initial magnetic field $B_{\rm i}$ (the initial spin period $P_{\rm i}$) distribution got from \citet{Faucher-Giguere2006} in the upper (lower) panel, respectively.
The black histogram lines in the upper and lower panel show the $B_{\rm i}$ and $P_{\rm i}$ got from our model, in which the solid, dashed and dot-dashed histogram lines in the lower panel represent the cases of $\xi=1$, $0.9$, $0.5$, respectively.
\label{fig:Bc}}
\end{figure}

We first consider the model without magnetic field decay and compare the results with the similar model from \citet{Faucher-Giguere2006} in Figure \ref{fig:Bc}, where the initial spin period is
\begin{equation}
    P_{\rm i} \approx (P^2-2C_1\times B_{\rm i}^2 \tau)^{1/2}.
    \label{equ:pi_bc}
\end{equation}
The grey histogram bars in the upper and lower panel indicate the surface dipole magnetic field $B_{\rm dip}$ and the observed spin period $P_{\rm obs}$ distribution of the NANSs, respectively.
%We find out that log-normal distributions can fit the data well, which follow $\langle {\rm log} (B_{\rm dip}/{\rm G}) \rangle \sim 12.06$, $\sigma_{\rm log ~ B_{\rm dip}} \sim 0.59$ in the upper panel and $\langle {\rm log} (P_{\rm obs}/{\rm s}) \rangle \sim -0.21$, $\sigma_{\rm log P_{\rm obs}} \sim 0.40$ in the lower panel.
The black histogram profile and the grey line indicate the initial magnetic field $B_{\rm i}$ and spin period $P_{\rm i}$ distribution got from our model and \citet{Faucher-Giguere2006} (FK 2006), respectively.
It shows that the distribution of $B_{\rm i}$ in our results is similar with that of \citet{Faucher-Giguere2006}, both of which follow log-normal distribution, while we obtain smaller mean value. 
However, $P_{\rm i}$ also pursues log-normal distribution in our model, which is $\langle {\rm log} (P_{\rm i}/{\rm s}) \rangle \sim -1.58$, $\sigma_{\rm log P_i} \sim 0.4$ for $\xi=1$. it is much different from the normal distribution of \citet{Faucher-Giguere2006}.
When we take $\xi<1$ \citep{Zhang2011}, i.e., we think the actual age of NSs is smaller than its spin-down age, the mean value of the $P_{\rm i}$ distribution becomes larger ($\langle {\rm log} (P_{\rm i}/{\rm s}) \rangle \sim -0.71$ for $\xi=0.9$ and $\langle {\rm log} (P_{\rm i}/{\rm s}) \rangle \sim -0.37$ for $\xi=0.5$) while the standard deviation does not change.
When we take $\xi>1$, we cannot get any value of $P_{\rm i}$.

\subsection{The model with magnetic field decay}
\label{sec:mdeol_Bt}

\begin{figure*}[ht!]
\plotone{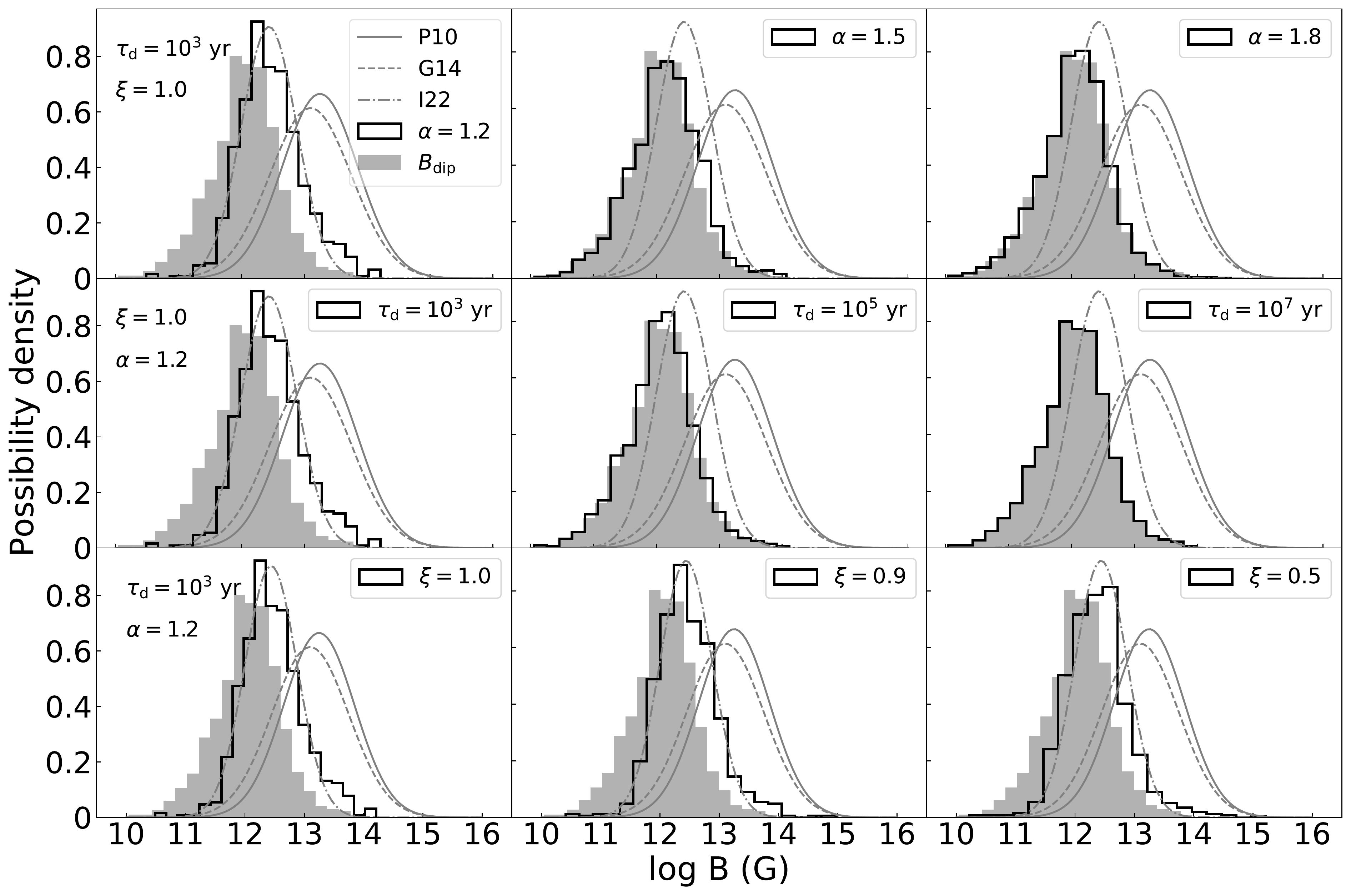}
\caption{The initial magnetic field distributions with different parameters. 
The parameters are set as follows.
Top panels: $\tau_{\rm d}=10^3 {\rm ~yr}$, $\xi=1.0$ and $\alpha = 1.2$, $1.5$, $1.8$ from left to right panel.
Middle panels: $\xi=1.0$,  $\alpha = 1.2$,  and $\tau_{\rm d}=10^3 {\rm ~yr}$, $10^5 {\rm ~yr}$, $10^7 {\rm ~yr}$ from left to right panel.
Lower panels: $\tau_{\rm d}=10^3 {\rm ~yr}$, $\alpha = 1.2$ and $\xi=1.0$, $0.9$, $0.5$ from left to right panel.
In comparison, the grey histogram bars display the distribution of NANSs' $B_{\rm dip}$ and the grey lines indicate the results from literature, where the solid, dashed and dot-dashed do that from \citep{Popov2010} (P10), \citep{Gullon2014} (G14) and \citep{Igoshev2022} (I22), respectively.
\label{fig:B_atx}}
\end{figure*}

Then we introduce magnetic field decay of NSs into the model and it (the left panel in Figure \ref{fig:Btau_pl}) shows that the power law form (Eq. \ref{equ:Bt}) is favoured.
We limit the parameters by fitting the distribution trend %\xiaotian{what is distribution tendency? I think this word is not accurate enough} 
of NANSs in $B_{\rm dip}-\tau_{\rm c}$ diagram (the lower panel in Figure \ref{fig:Btau_pl}) in a rough range: $1.2 \leq \alpha \leq 1.8$ and $10^3 \leq \tau_{\rm d}/{\rm yr} \leq 10^7$.
Then we try to simulate the initial values of NANSs' magnetic field and spin period within the above parameter intervals.

Figure \ref{fig:B_atx} displays the results of initial magnetic field distributions obtained by  the control variable method. 
The variables are $\alpha$, $\tau_{\rm d}$ and $\xi$ from top to bottom panels and other parameters are set to be $\alpha=1.2$, $\tau_{\rm d}=10^3 {\rm ~yr}$ and $\xi=1.0$ when they are not variables.
It shows that all $B_{\rm i}$ in our model can be fitting well with log-normal distributions with $\langle {\rm log} (B_{\rm i}/{\rm G}) \rangle \sim 12.04-12.47$ and $\sigma _{\rm log ~ B_i} \sim 0.5-0.6$.
In the case with $\alpha=1.2$, $\tau_{\rm d}=10^3 {\rm ~yr}$ and $\xi=1.0$, we get $\langle {\rm log} (B_{\rm i}/{\rm G}) \rangle \sim 12.47$, $\sigma _{\rm log ~ B_i} \sim 0.5$, which is consistent with the last results in \citet{Igoshev2022}.

\begin{figure*}[ht!]
\plotone{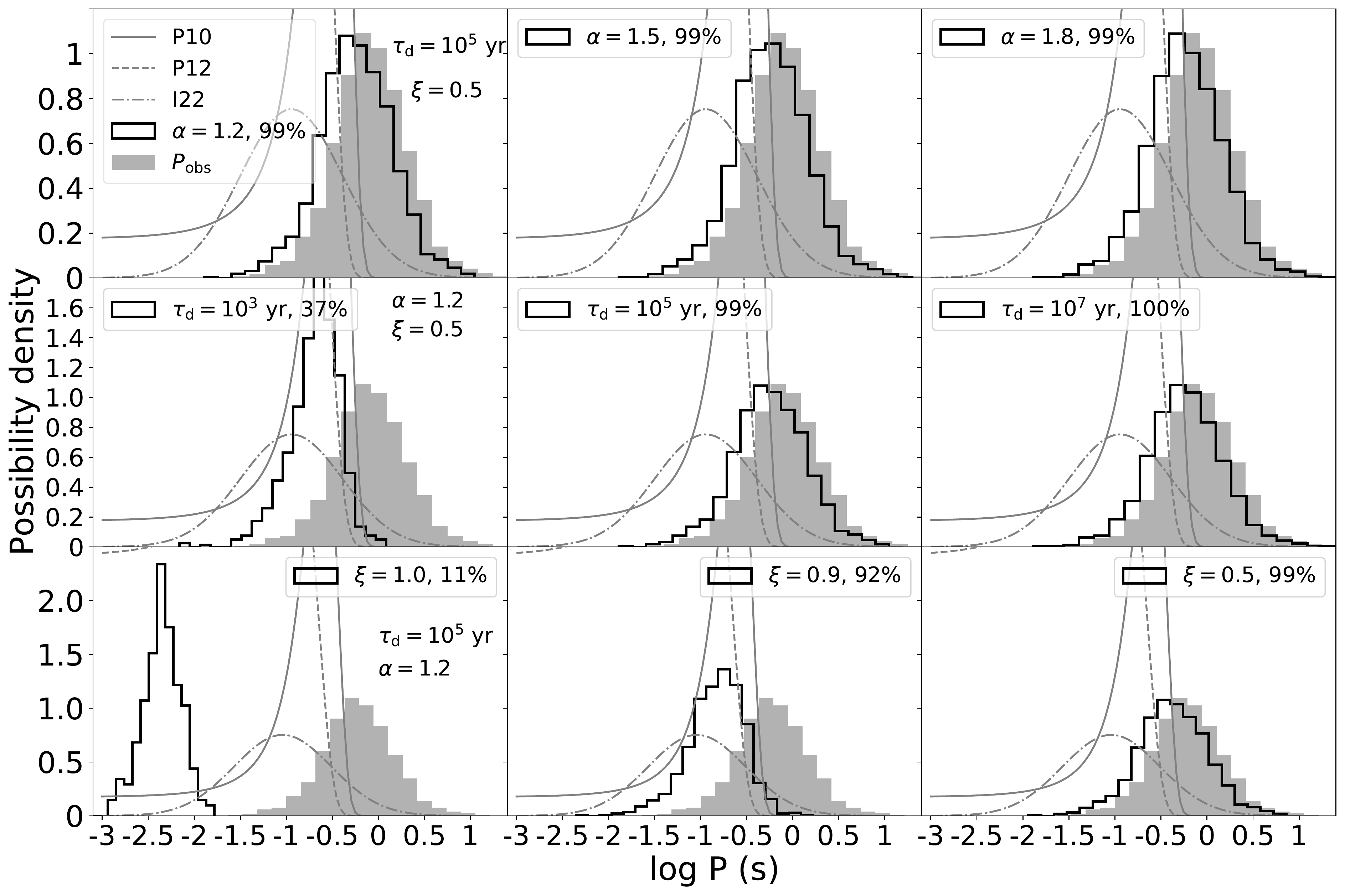}
\caption{The initial spin period distributions with different parameters. 
The parameters are set as follows.
Top panels: $\tau_{\rm d}=10^5 {\rm ~yr}$, $\xi=0.5$ and $\alpha = 1.2$, $1.5$, $1.8$ from left to right panel.
Middle panels: $\alpha = 1.2$, $\xi=0.5$, and $\tau_{\rm d}=10^3 {\rm ~yr}$, $10^5 {\rm ~yr}$, $10^7 {\rm ~yr}$ from left to right panel.
Lower panels: $\tau_{\rm d}=10^5 {\rm ~yr}$, $\alpha = 1.2$ and $\xi=1.0$, $0.9$, $0.5$ from left to right panel.
In contrast, the gray histogram bars display the distribution of NANSs' $P_{\rm obs}$ and the grey lines indicate the results from literature, where the solid, dashed and dot-dashed do that from \citep{Popov2010} (P10), \citep{Popov2012} (P12) and \citep{Igoshev2022} (I22), respectively.
The percentage numbers in the labels indicate the proportion of valid samples to total samples, which is ratio of the size of the sources that we can get its $P_{\rm i}$ to the NANSs.
\label{fig:P_atx}}
\end{figure*}

When using the same parameters to simulate the spin periods, we can only get the initial values for part of NANSs.
Figure \ref{fig:P_atx} shows the initial spin period distributions under the variable of $\alpha$, $\tau_{\rm d}$ and $\xi$ from top to bottom, where the non-variable parameters are set to be $\alpha=1.2$, $\tau_{\rm d}=10^5 {\rm ~yr}$ and $\xi=0.5$.
The percentage numbers in the figure labels indicate the proportion of valid sample, %\xiaotian{valid samples?}, 
which are the NANSs that we can get its initial spin periods,
%\xiaotian{This gives me an impression that you are using different samples for magnetic field calculations and spin period calculations}, 
to total sample, which are all the NANSs.
It shows that all the $P_{\rm i}$ histograms favour log-normal distributions with $\langle {\rm log} (P_{\rm i}/{\rm s}) \rangle \sim (-2.37) - (-0.37)$ and $\sigma _{\rm log ~ P_i} \sim 0.2-0.4$ whatever full or partial samples, in which the 6 upper right panels (the upper three, the middle-middle, the middle-right and the lower right panels) share the same results of $\langle {\rm log} (P_{\rm i}/{\rm s}) \rangle \sim -0.37$ and $\sigma _{\rm log ~ P_i} \sim 0.4$.

%\xiaotian{it would be nice if you make a table to summarize your results given by different input physics}

\section{Discussion and Summary} \label{sec:discussion_summary}

%\xiaotian{after reading this section, I can not figure out what is your key message.}

%\xiaotian{what is the key certainty of this work? What happens if some NSs of your sample are actually binary product?}

%\xiaotian{I recall that some works proposed that NSs have their B field rapidly decaying from $10^{15}$ G to $10^{12}$ G within their first $10^5$ years. Is this process considered?}

In this work we simulate the initial distribution of magnetic fields $B_{\rm i}$ and spin periods $P_{\rm i}$ of non-accretion neutron stars, which are thought to have only undergone magnetic dipole radiation.
Our result shows that both $B_{\rm i}$ and $P_{\rm i}$ favour log-normal distributions, 
while their values depend strongly on parameters.
%while their values are very affcted by parameters.

Since the relation $\xi=\tau/\tau_{\rm c}$ between the physical age $\tau$ and spin-down age $\tau_{\rm c}$ of NSs is unknown, we take a linear one for simplicity.
It shows that the mean value of $B_{\rm i}$ increases but that of $P_{\rm i}$ decreases as $\xi$ becomes larger (the lower panels in Figure \ref{fig:B_atx} and \ref{fig:P_atx}). 
Furthermore, $\xi$ has little effect on the mean value of $B_{\rm i}$ (the lower panels in Figure \ref{fig:B_atx}) but big on $P_{\rm i}$ (the lower panels in Figure \ref{fig:Bc} and Figure \ref{fig:P_atx}).
Even in some situations, the simulation cannot give the $P_{\rm i}$ values for some sources (Figure \ref{fig:P_atx}). it is because the integration term in Eq. \ref{equ:p} may be larger than the square term of the observed spin period, i.e., there may be $2C_1\cdot \int_{t_0}^{t} B^2(t)\ dt > P^2$ in some cases if the age in the integration is longer than the physical age of pulsars.
So we think the unknown relation should meet the condition $\xi \leq 1$ \citep{Zhang2011}.
The upper panels in Figure \ref{fig:B_atx} and \ref{fig:P_atx} indicate that the mean value of $B_{\rm i}$ increases as the parameter $\alpha$ becomes smaller while there is nearly no difference among the histogram lines of $P_{\rm i}$.
The effect of the parameter $\tau_{\rm d}$ is opposite to  $\xi$, i.e., the mean value of $B_{\rm i}$ decreases but that of $P_{\rm i}$ increases as $\tau_{\rm d}$ becomes larger (the middle panels in Figure \ref{fig:B_atx} and \ref{fig:P_atx}).

The following arguments may lead to the initial distributions depending strongly on parameters in our simulation: 
\begin{enumerate}
    \item The non-accretion neutron stars we picked are probably not representative of the pulsars which have only undergone magnetic dipole radiation. 
    In fact, most of pulsars may experience other processes of losing angular momentum, such as glitches \citep{Espinoza2011}, accretion \citep{Ghosh1979} or magnetised stellar wind \citep{Bisnovatyi-Kogan2017}.
    \item The decay forms and parameters of magnetic field may be various for different NSs since the nature and characteristics are diverse, such as the mass \citep{Kiziltan2013}, the initial magnetic field, the equation of state and the internal structure \citep{Sumiyoshi2022}.
    \item The relation between the physical and spin-down age of NSs may be more complex than a linear one \citep{Camilo1994}.
\end{enumerate}
However, since our sample is large enough and the physical age of NSs is too hard to detect, so our result is still somewhat informative.

\begin{acknowledgments}
We thank the anonymous referee for helpful comments and suggestions.
This work was supported by the Natural Natural
Science Foundation of China (NSFC, Grant Nos. 11988101, 12203051, 11933004, 12041301, 12063001, 11773015), Project funded by China Postdoctoral Science Foundation No. 2021M703168.
\end{acknowledgments}

%\clearpage
%\newpage
\bibliography{sample631}{}

\begin{thebibliography}{}
\expandafter\ifx\csname natexlab\endcsname\relax\def\natexlab#1{#1}\fi
\providecommand{\url}[1]{\href{#1}{#1}}
\providecommand{\dodoi}[1]{doi:~\href{http://doi.org/#1}{\nolinkurl{#1}}}
\providecommand{\doeprint}[1]{\href{http://ascl.net/#1}{\nolinkurl{http://ascl.net/#1}}}
\providecommand{\doarXiv}[1]{\href{https://arxiv.org/abs/#1}{\nolinkurl{https://arxiv.org/abs/#1}}}

\bibitem[{{Aguilera} {et~al.}(2008){Aguilera}, {Pons}, \&
  {Miralles}}]{Aguilera2008}
{Aguilera}, D.~N., {Pons}, J.~A., \& {Miralles}, J.~A. 2008, \aap, 486, 255,
  \dodoi{10.1051/0004-6361:20078786}

\bibitem[{{Bisnovatyi-Kogan}(2017)}]{Bisnovatyi-Kogan2017}
{Bisnovatyi-Kogan}, G.~S. 2017, in Handbook of Supernovae, ed. A.~W. {Alsabti}
  \& P.~{Murdin}, 1401, \dodoi{10.1007/978-3-319-21846-5_70}

\bibitem[{{Camilo} {et~al.}(1994){Camilo}, {Thorsett}, \&
  {Kulkarni}}]{Camilo1994}
{Camilo}, F., {Thorsett}, S.~E., \& {Kulkarni}, S.~R. 1994, \apjl, 421, L15,
  \dodoi{10.1086/187176}

\bibitem[{{Colpi} {et~al.}(2000){Colpi}, {Geppert}, \& {Page}}]{Colpi2000}
{Colpi}, M., {Geppert}, U., \& {Page}, D. 2000, \apjl, 529, L29,
  \dodoi{10.1086/312448}

\bibitem[{{Dall'Osso} {et~al.}(2012){Dall'Osso}, {Granot}, \&
  {Piran}}]{Dall'Osso2012}
{Dall'Osso}, S., {Granot}, J., \& {Piran}, T. 2012, \mnras, 422, 2878,
  \dodoi{10.1111/j.1365-2966.2012.20612.x}

\bibitem[{{De Luca} {et~al.}(2006){De Luca}, {Caraveo}, {Mereghetti}, {Tiengo},
  \& {Bignami}}]{DeLuca2006}
{De Luca}, A., {Caraveo}, P.~A., {Mereghetti}, S., {Tiengo}, A., \& {Bignami},
  G.~F. 2006, Science, 313, 814, \dodoi{10.1126/science.1129185}

\bibitem[{{Espinoza} {et~al.}(2011){Espinoza}, {Lyne}, {Stappers}, \&
  {Kramer}}]{Espinoza2011}
{Espinoza}, C.~M., {Lyne}, A.~G., {Stappers}, B.~W., \& {Kramer}, M. 2011,
  \mnras, 414, 1679, \dodoi{10.1111/j.1365-2966.2011.18503.x}

\bibitem[{{Faucher-Gigu{\`e}re} \& {Kaspi}(2006)}]{Faucher-Giguere2006}
{Faucher-Gigu{\`e}re}, C.-A., \& {Kaspi}, V.~M. 2006, \apj, 643, 332,
  \dodoi{10.1086/501516}

\bibitem[{{Fu} \& {Li}(2012)}]{Fu2012}
{Fu}, L., \& {Li}, X.-D. 2012, \apj, 757, 171,
  \dodoi{10.1088/0004-637X/757/2/171}

\bibitem[{{Ghosh} \& {Lamb}(1979)}]{Ghosh1979}
{Ghosh}, P., \& {Lamb}, F.~K. 1979, \apj, 234, 296, \dodoi{10.1086/157498}

\bibitem[{{Goldreich} \& {Reisenegger}(1992)}]{Goldreich1992}
{Goldreich}, P., \& {Reisenegger}, A. 1992, \apj, 395, 250,
  \dodoi{10.1086/171646}

\bibitem[{{Gonthier} {et~al.}(2002){Gonthier}, {Ouellette}, {Berrier},
  {O'Brien}, \& {Harding}}]{Gonthier2002}
{Gonthier}, P.~L., {Ouellette}, M.~S., {Berrier}, J., {O'Brien}, S., \&
  {Harding}, A.~K. 2002, \apj, 565, 482, \dodoi{10.1086/324535}

\bibitem[{{Gonthier} {et~al.}(2004){Gonthier}, {Van Guilder}, \&
  {Harding}}]{Gonthier2004}
{Gonthier}, P.~L., {Van Guilder}, R., \& {Harding}, A.~K. 2004, \apj, 604, 775,
  \dodoi{10.1086/382070}

\bibitem[{{Gull{\'o}n} {et~al.}(2014){Gull{\'o}n}, {Miralles}, {Vigan{\`o}}, \&
  {Pons}}]{Gullon2014}
{Gull{\'o}n}, M., {Miralles}, J.~A., {Vigan{\`o}}, D., \& {Pons}, J.~A. 2014,
  \mnras, 443, 1891, \dodoi{10.1093/mnras/stu1253}

\bibitem[{{Igoshev} {et~al.}(2022){Igoshev}, {Frantsuzova}, {Gourgouliatos},
  {Tsichli}, {Konstantinou}, \& {Popov}}]{Igoshev2022}
{Igoshev}, A.~P., {Frantsuzova}, A., {Gourgouliatos}, K.~N., {et~al.} 2022,
  \mnras, 514, 4606, \dodoi{10.1093/mnras/stac1648}

\bibitem[{{Igoshev} \& {Popov}(2013)}]{Igoshev2013}
{Igoshev}, A.~P., \& {Popov}, S.~B. 2013, \mnras, 432, 967,
  \dodoi{10.1093/mnras/stt519}

\bibitem[{{Igoshev} {et~al.}(2021){Igoshev}, {Popov}, \&
  {Hollerbach}}]{Igoshev2021}
{Igoshev}, A.~P., {Popov}, S.~B., \& {Hollerbach}, R. 2021, Universe, 7, 351,
  \dodoi{10.3390/universe7090351}

\bibitem[{{Jawor} \& {Tauris}(2022)}]{Jawor2022}
{Jawor}, J.~A., \& {Tauris}, T.~M. 2022, \mnras, 509, 634,
  \dodoi{10.1093/mnras/stab2677}

\bibitem[{{Kiel} {et~al.}(2008){Kiel}, {Hurley}, {Bailes}, \&
  {Murray}}]{Kiel2008}
{Kiel}, P.~D., {Hurley}, J.~R., {Bailes}, M., \& {Murray}, J.~R. 2008, \mnras,
  388, 393, \dodoi{10.1111/j.1365-2966.2008.13402.x}

\bibitem[{{Kiziltan} {et~al.}(2013){Kiziltan}, {Kottas}, {De Yoreo}, \&
  {Thorsett}}]{Kiziltan2013}
{Kiziltan}, B., {Kottas}, A., {De Yoreo}, M., \& {Thorsett}, S.~E. 2013, \apj,
  778, 66, \dodoi{10.1088/0004-637X/778/1/66}

\bibitem[{{Konar}(2017)}]{Konar2017}
{Konar}, S. 2017, Journal of Astrophysics and Astronomy, 38, 47,
  \dodoi{10.1007/s12036-017-9467-4}

\bibitem[{{Livio} {et~al.}(1998){Livio}, {Xu}, \& {Frank}}]{Livio1998}
{Livio}, M., {Xu}, C., \& {Frank}, J. 1998, \apj, 492, 298,
  \dodoi{10.1086/305034}

\bibitem[{{Makarenko} {et~al.}(2021){Makarenko}, {Igoshev}, \&
  {Kholtygin}}]{Makarenko2021}
{Makarenko}, E.~I., {Igoshev}, A.~P., \& {Kholtygin}, A.~F. 2021, \mnras, 504,
  5813, \dodoi{10.1093/mnras/stab1175}

\bibitem[{{Manchester} {et~al.}(2005){Manchester}, {Hobbs}, {Teoh}, \&
  {Hobbs}}]{Manchester2005}
{Manchester}, R.~N., {Hobbs}, G.~B., {Teoh}, A., \& {Hobbs}, M. 2005, \aj, 129,
  1993, \dodoi{10.1086/428488}

\bibitem[{{Mendes} {et~al.}(2018){Mendes}, {de Avellar}, {Horvath}, {Souza},
  {Benvenuto}, \& {De Vito}}]{Mendes2018}
{Mendes}, C., {de Avellar}, M. G.~B., {Horvath}, J.~E., {et~al.} 2018, \mnras,
  475, 2178, \dodoi{10.1093/mnras/stx3319}

\bibitem[{{Olausen} \& {Kaspi}(2014)}]{Olausen2014}
{Olausen}, S.~A., \& {Kaspi}, V.~M. 2014, \apjs, 212, 6,
  \dodoi{10.1088/0067-0049/212/1/6}

\bibitem[{{Os{\l}owski} {et~al.}(2011){Os{\l}owski}, {Bulik},
  {Gondek-Rosi{\'n}ska}, \& {Belczy{\'n}ski}}]{Oslowski2011}
{Os{\l}owski}, S., {Bulik}, T., {Gondek-Rosi{\'n}ska}, D., \& {Belczy{\'n}ski},
  K. 2011, \mnras, 413, 461, \dodoi{10.1111/j.1365-2966.2010.18147.x}

\bibitem[{{Pons} \& {Geppert}(2007)}]{Pons2007}
{Pons}, J.~A., \& {Geppert}, U. 2007, \aap, 470, 303,
  \dodoi{10.1051/0004-6361:20077456}

\bibitem[{{Popov} {et~al.}(2010){Popov}, {Pons}, {Miralles}, {Boldin}, \&
  {Posselt}}]{Popov2010}
{Popov}, S.~B., {Pons}, J.~A., {Miralles}, J.~A., {Boldin}, P.~A., \&
  {Posselt}, B. 2010, \mnras, 401, 2675,
  \dodoi{10.1111/j.1365-2966.2009.15850.x}

\bibitem[{{Popov} \& {Turolla}(2012)}]{Popov2012}
{Popov}, S.~B., \& {Turolla}, R. 2012, \apss, 341, 457,
  \dodoi{10.1007/s10509-012-1100-z}

\bibitem[{{Sumiyoshi} {et~al.}(2022){Sumiyoshi}, {Kojo}, \&
  {Furusawa}}]{Sumiyoshi2022}
{Sumiyoshi}, K., {Kojo}, T., \& {Furusawa}, S. 2022, arXiv e-prints,
  arXiv:2207.00033.
\newblock \doarXiv{2207.00033}

\bibitem[{{Usov}(1988)}]{Usov1988}
{Usov}, V.~V. 1988, \apss, 140, 39, \dodoi{10.1007/BF00643526}

\bibitem[{{Xu} \& {Li}(2019)}]{Xu2019}
{Xu}, K., \& {Li}, X.-D. 2019, \apj, 877, 138, \dodoi{10.3847/1538-4357/ab1902}

\bibitem[{{Xu} {et~al.}(2022){Xu}, {Li}, {Cui}, {Li}, {Shao}, {Liang}, \&
  {Liu}}]{Xu2022}
{Xu}, K., {Li}, X.-D., {Cui}, Z., {et~al.} 2022, Research in Astronomy and
  Astrophysics, 22, 015005, \dodoi{10.1088/1674-4527/ac321f}

\bibitem[{{Zhang} \& {Xie}(2011)}]{Zhang2011}
{Zhang}, S., \& {Xie}, Y. 2011, in Astronomical Society of the Pacific
  Conference Series, Vol. 451, 9th Pacific Rim Conference on Stellar
  Astrophysics, ed. S.~{Qain}, K.~{Leung}, L.~{Zhu}, \& S.~{Kwok}, 231.
\newblock \doarXiv{1110.3154}

\bibitem[{{Zhang} \& {Xie}(2012)}]{Zhang2012}
{Zhang}, S.-N., \& {Xie}, Y. 2012, \apj, 757, 153,
  \dodoi{10.1088/0004-637X/757/2/153}

\end{thebibliography}
\bibliographystyle{aasjournal}

\end{document}